\newif\ifarxiv

\arxivfalse
\arxivtrue

\documentclass[journal,twoside,web]{ieeecolor}
\usepackage{lcsys}

\usepackage{graphicx}
\usepackage{amssymb}
\usepackage{amsmath} 
\usepackage{color}
\usepackage[algo2e,vlined]{algorithm2e}
\usepackage[hidelinks]{hyperref}

\newcommand{\R}{\mathbb{R}}

\newtheorem{problem}{Problem}
\newtheorem{proposition}[problem]{Proposition}
\newtheorem{theorem}[problem]{Theorem}
\newtheorem{example}{Example}

\graphicspath{{./Images/}}

\begin{document}

\title{Reachability analysis of neural networks using mixed monotonicity}

\author{Pierre-Jean Meyer
\thanks{Pierre-Jean Meyer is with Univ Gustave Eiffel, COSYS-ESTAS, F-59666 Villeneuve d'Ascq, France
{\tt\small pierre-jean.meyer@univ-eiffel.fr}}%
}

\maketitle
\thispagestyle{empty}
\pagestyle{empty}

\begin{abstract}
This paper presents a new reachability analysis approach to compute interval over-approximations of the output set of feedforward neural networks with input uncertainty.
We adapt to neural networks an existing mixed-monotonicity method for the reachability analysis of dynamical systems and apply it to each partial network within the main network.
This ensures that the intersection of the obtained results is the tightest interval over-approximation of the output of each layer that can be obtained using mixed-monotonicity on any partial network decomposition.
Unlike other tools in the literature focusing on small classes of piecewise-affine or monotone activation functions, the main strength of our approach is its generality: it can handle neural networks with any Lipschitz-continuous activation function.
In addition, the simplicity of our framework allows users to very easily add unimplemented activation functions, by simply providing the function, its derivative and the global argmin and argmax of the derivative.
Our algorithm is compared to five other interval-based tools (Interval Bound Propagation, \textsc{ReluVal}, \textsc{Neurify}, \textsc{VeriNet}, CROWN) on both existing benchmarks and two sets of small and large randomly generated networks for four activation functions (ReLU, TanH, ELU, SiLU).
\end{abstract}


\section{Introduction}
\label{sec intro}
\IEEEPARstart{A}{rtificial}
intelligence and particularly neural networks are quickly spreading in various fields including safety-critical applications such as autonomous driving~\cite{xiang2018verification}.
With it comes a growing need for replacing verification methods based on statistical testing~\cite{kim2020programmatic} by more formal methods able to provide safety guarantees of the satisfaction of desired properties on the output of the neural network~\cite{liu2019algorithms}.
Such methods are also useful for the verification of a closed-loop system with a neural network controller~\cite{akintunde2018reachability,xiang2020reachable}.

When focusing on isolated neural networks, the rapidly growing field of formal verification has been categorized into three main types of objectives~\cite{liu2019algorithms}: \emph{counter-example result} to find an input-output pair that violates the desired property; \emph{adversarial result} to determine the maximum allowed disturbance that can be applied to a nominal input while preserving the properties of the nominal output; \emph{reachability result} to evaluate (exactly or through over-approximations) the set of all output values reachable by the network when given a bounded input set.
While the first two types of results often rely on solving optimization problems~\cite{liu2019algorithms,salman2019convex}, reachability results (which are considered in this paper) naturally lean toward developing new reachability analysis approaches to be applied layer by layer in the network.
The current reachability methods in the literature consider various set representations to over-approximate the network's output set with static intervals~\cite{xiang2020reachable}, symbolic interval equations and linear relaxations of activation functions (\textsc{ReluVal}~\cite{shiqi2018reluval}, \textsc{Neurify}~\cite{shiqi2018neurify}, \textsc{VeriNet}~\cite{henriksen2020efficient}, CROWN~\cite{zhang2018efficient}) or polyhedra and zonotopes in the tool libraries NNV~\cite{tran2020nnv} and ERAN~\cite{eran}.

The main weakness of existing neural network verifiers is that most of them are limited to a very small class of activation functions.
The large majority of verifiers only handles piecewise-affine activation to focus on the most popular ReLU function~\cite{shiqi2018reluval,shiqi2018neurify,katz2019marabou,botoeva2020efficient}.
A handful of tools cover other popular activations, such as S-shaped functions sigmoid and hyperbolic tangent~\cite{henriksen2020efficient,tran2020nnv,eran}.
Several other tools claiming to handle general activation functions are either implicitly restricted to monotone functions in their theory or implementation~\cite{dvijotham2018dual,raghunathan2018certified}, or their claimed generality rather refers to a mere compatibility with general activation functions but they require the user to provide their own implementation of how to handle these new functions~\cite{zhang2018efficient}.

Although ReLU or sigmoid networks often perform well, the focus on these activation functions in the literature is also strongly tied to the fact that most current optimization and verification tools cannot handle more general activations. Proposing a new approach to deal with these general activation functions is thus an important step towards opening the field of formal verification of neural networks to new non-monotone activations that have already been shown to outperform classical ReLU or sigmoid networks (see e.g.\ Swish/SiLU~\cite{ramachandran2017searching}, Mish~\cite{misra2019mish}, PFLU and FPFLU~\cite{zhu2021pflu}).

\paragraph*{Contributions}
This paper presents a novel method for the reachability analysis of feedforward neural networks with general activation functions.
The proposed approach adapts the existing mixed-monotonicity reachability method for dynamical systems~\cite{meyer2021springer} to be applicable to neural networks so that we can obtain an interval over-approximation of the set of all the network's outputs that can be reached from a given input set.
Since our reachability method is also applicable to any partial network within the main neural network and always returns a sound over-approximation of the last layer's output set, intersecting the interval over-approximations from several partial networks ending at the same layer can only tighten the approximation while preserving the soundness of the results.
To take full advantage of this, we propose an algorithm that applies our new mixed-monotonicity reachability method to all $L(L+1)/2$ partial networks contained within the considered $L$-layer network.
This ensures that the resulting interval over-approximation is the tightest output bound of the network obtainable using mixed-monotonicity reachability analysis on any partial network decomposition, at the cost of a computational complexity of $O(L^2)$.

Mixed-monotonicity reachability analysis is applicable to any system whose Jacobian matrix is bounded on any bounded input set~\cite{meyer2021springer}.
In the case of neural networks combining linear transformations and nonlinear activation functions, the above requirement implies that our approach is applicable to any network with Lipschitz-continuous activation function.
Since extracting the bounds of the Jacobian matrix of a system (or of the derivative of an activation function in our case) is not always straightforward, for the sake of self-containment of the paper we provide a method to automatically obtain these bounds for a (still very general) sub-class of activation functions whose derivative can be defined as a $3$-piece piecewise-monotone function.
\ifarxiv
Apart from the binary step and Gaussian function, all activation functions we could find in the literature (including all non-monotone functions reviewed in~\cite{zhu2021pflu}) belong to this class.
\fi

Although most of the tools cited above provide a two-part verification framework for neural networks (one reachability part to compute output bounds of the network, and one part to iteratively refine the network's domain until a definitive answer can be given to the verification problem), this paper focuses on providing a novel method for the first reachability step only.
The proposed approach can thus be seen either as a preliminary step towards the development of a larger verification framework for neural networks when combined with an iterative splitting of the input domain, or as a stand-alone tool for the reachability analysis of neural networks as is used for example in the analysis of closed-loop systems with a sample-and-hold neural network controller~\cite{xiang2020reachable}.

In summary, the main contributions of this paper are:
\begin{itemize}
\item a novel approach to soundly bound the output of neural networks using mixed-monotonicity reachability;
\item a method that is compatible with any Lipschitz-continuous activation function;
\item a framework that allows to easily add new activation functions (the user only needs to provide the activation function, its derivative, and the global $\arg\min$ and $\arg\max$ of the derivative).
\end{itemize}

The paper is organized as follows.
Section~\ref{sec prelim} defines the considered problem and provides useful preliminaries for the main algorithm, such as the definition of mixed-monotonicity reachability for a neural network, and how to automatically obtain local bounds on the activation function derivative.
Section~\ref{sec reachability} presents the main algorithm applying mixed-monotonicity reachability to all partial networks within the given network.
Finally, Section~\ref{sec numerical} compares our novel reachability approach to the bounding methods of $5$ other interval-based optimization-free tools~\cite{xiang2020reachable,shiqi2018reluval,shiqi2018neurify,henriksen2020efficient,zhang2018efficient} on both existing benchmarks~\cite{bak2021second} and randomly generated networks, to highlight the complementarity with these tools and the cases when our approach outperforms them.

\section{Preliminaries}
\label{sec prelim}
Given $\underline{x},\overline{x}\in\R^n$ with $\underline{x}\leq\overline{x}$, the interval $[\underline{x},\overline{x}]\subseteq\R^n$ is the set $\{x\in\R^n~|~\forall i\in\{1,\dots,n\},~\underline{x}_i\leq x_i\leq\overline{x}_i\}$.

\subsection{Problem definition}
\label{sub prelim pb}
Consider the $L$-layer feedforward neural network 
\begin{equation}
\label{eq ffnn}
x^{(i)}=\Phi(W^{(i)}x^{(i-1)}+b^{(i)}),~\forall i\in\{1,\dots,L\}
\end{equation}
where $x^{(i)}\in\R^{n_i}$ is the output vector of layer $i$, $x^{(0)}$ and $x^{(L)}$ being the input and output of the neural network, respectively.
We assume that the network is pre-trained and all weight matrices $W^{(i)}\in\R^{n_{i}\times n_{i-1}}$ and bias vectors $b^{(i)}\in\R^{n_{i}}$ are known.
The function $\Phi$ is defined as the componentwise application of a scalar and Lipschitz-continuous activation function.
For simplicity of presentation, the activation function $\Phi$ is assumed to be identical for all layers, but note that the proposed approach in this paper is compatible with having different activation functions for each layer (or even for each node of the same layer).
In the literature, the activation function of the last layer is often omitted since a monotone activation (the most commonly used) would only change the output values but not their relative comparison (which is considered for classification problems).
Since our approach introduces the ability to handle non-monotone activation, our network definition \eqref{eq ffnn} needs to allow for the activation of all layers.

Some verification problems on neural networks aim to measure the robustness of the network with respect to input variations.
Since the output set of the network cannot always be computed exactly, we rely on over-approximating it by a simpler set representation, such as an interval.
We can then solve the verification problem using this interval: if the desired property is satisfied on the over-approximation, then it is also satisfied on the real output set of the network.
In this paper, we focus on the problem of computing an interval over-approximation of the output set of the network when its input is taken in a known bounded set, as formalized below.
\begin{problem}
\label{pb definition}
Given the $L$-layer neural network~\eqref{eq ffnn} and the interval input set $[\underline{x^0},\overline{x^0}]\subseteq\R^{n_0}$, find an interval $[\underline{x^L},\overline{x^L}]\subseteq\R^{n_L}$ over-approximating the output set of \eqref{eq ffnn}:
$$\{x^{(L)}~|~x^{(0)}\in[\underline{x^0},\overline{x^0}]\}\subseteq[\underline{x^L},\overline{x^L}].$$
\end{problem}

Naturally, our secondary objective is to ensure that the computed interval over-approximation is as tight as possible in order to minimize the number of false negative results in the subsequent verification process.

\subsection{Mixed-monotonicity reachability}
\label{sub prelim MM}
In this paper, we solve Problem~\ref{pb definition} by iteratively computing over-approximation intervals of the output of each layer, until the last layer of the network is reached.
These over-approximations are obtained by adapting to neural networks existing methods for reachability analysis of dynamical systems.
More specifically, we rely on the mixed-monotonicity approach in~\cite{meyer2021springer} to over-approximate the reachable set of any discrete-time system $x^+=f(x)$ with Lipschitz-continuous vector field $f:\R^n\rightarrow\R^n$.
Since the neural network \eqref{eq ffnn} is instead defined as a function $y=f(x)$ with input $x=x^{(0)}$ and output $y=x^{(L)}$ of different dimensions, we present below the generalization of the mixed-monotonicity method from~\cite{meyer2021springer} to any Lipschitz-continuous function.

Consider the function $y=f(x)$ with input $x\in[\underline{x},\overline{x}]\subseteq\R^{n_x}$, output $y\in\R^{n_y}$ and Lipschitz-continuous function $f:\R^{n_x}\rightarrow\R^{n_y}$.
The Lipschitz-continuity assumption ensures that the derivative $f'$ (also called Jacobian matrix in the paper) is bounded.
\begin{proposition}
\label{prop algebraic MM}
Given an interval $[\underline{J},\overline{J}]\subseteq\R^{n_y\times n_x}$ bounding the derivative $f'(x)$ for all $x\in[\underline{x},\overline{x}]$, denote the center of the interval as $J^*$.
For each output dimension $i\in\{1,\dots,n_y\}$, define input vectors $\underline{\xi^i},\overline{\xi^i}\in\R^{n_x}$ and row vector $\alpha^i\in\R^{1\times n_x}$ such that for all $j\in\{1,\dots,n_x\}$,
$$(\underline{\xi^i}_j,\overline{\xi^i}_j,\alpha^i_j)=
\begin{cases}
(\underline{x}_j,\overline{x}_j,\min(0,\underline{J}_{ij}))&\text{if }J^*_{ij}\geq0,\\
(\overline{x}_j,\underline{x}_j,\max(0,\overline{J}_{ij}))&\text{if }J^*_{ij}\leq0.
\end{cases}$$
Then for all $x\in[\underline{x},\overline{x}]$ and $i\in\{1,\dots,n_y\}$, we have:
$$f_i(x)\in\left[f_i(\underline{\xi^i})-\alpha^i(\underline{\xi^i}-\overline{\xi^i}),f_i(\overline{\xi^i})+\alpha^i(\underline{\xi^i}-\overline{\xi^i})\right].$$
\end{proposition}

Intuitively, the output bounds are obtained by computing for each output dimension the images for two diagonally opposite vertices of the input interval, then expanding these bounds with an error term when the bounds on the derivative $f'$ spans both negative and positive values.
Proposition~\ref{prop algebraic MM} can thus provide an interval over-approximation of the output set of any function as long as bounds on the derivative $f'$ are known.
Obtaining such bounds for a neural network is made possible by computing local bounds on the derivative of its activation functions, as detailed in Section~\ref{sub prelim af bounds}.

\subsection{Local bounds of activation functions}
\label{sub prelim af bounds}
Proposition~\ref{prop algebraic MM} can be applied to any neural network whose activation functions are Lipschitz continuous since such functions have a bounded derivative.
However, to avoid requiring the users to manually provide these bounds for each different activation function they want to use, we instead provide a framework to automatically and easily compute local bounds for a very general class of functions describing most popular activation functions and their derivatives.

Let $\R_\infty=\R\cup\{-\infty,+\infty\}$ and consider the scalar function $\varphi:\R_\infty\rightarrow\R_\infty$, where $\varphi(x)\in\{-\infty,+\infty\}$ only when $x\in\{-\infty,+\infty\}$.
In this paper, we focus on $3$-piece piecewise-monotone functions for which there exist $\underline{z},\overline{z}\in\R_\infty$ such that $\varphi$ is:
\begin{itemize}
\item non-increasing on $(-\infty,\underline{z}]$ until reaching its global minimum $\min_{x\in\R_\infty}\varphi(x)=\varphi(\underline{z})$;
\item non-decreasing on $[\underline{z},\overline{z}]$ until reaching its global maximum $\max_{x\in\R_\infty}\varphi(x)=\varphi(\overline{z})$; 
\item and non-increasing on $[\overline{z},+\infty)$.
\end{itemize}
When $\underline{z}=-\infty$ (resp.\ $\overline{z}=+\infty$), the first (resp.\ last) monotone segment is squeezed into a singleton at infinity and can thus be ignored.

Although this description may appear restrictive, we should note that the vast majority of activation functions as well as their derivatives belong to this class of functions.
In particular, this is the case for (but not restricted to) popular piecewise-affine activation functions (identity, ReLU, leaky ReLU), monotone functions (hyperbolic tangent, arctangent, sigmoid, SoftPlus, ELU) as well as many non-monotone activation functions (SiLU, GELU, HardSwish, Mish, REU, PFLU, FPFLU, which are all reviewed or introduced in~\cite{zhu2021pflu}).
\ifarxiv
\begin{example}
Two examples of such functions are provided in Figure~\ref{fig silu}.
The first one (in black) is the Sigmoid Linear Unit (SiLU, also called \emph{Swish}~\cite{ramachandran2017searching}) activation function $\Phi(x)=x/(1+e^{-x})$, whose global $\arg\min$ is $\underline{z}=-1.2785$.
Its global $\arg\max$ is $\overline{z}=+\infty$, which implies that the third monotone component is pushed toward $+\infty$ since it is not needed for this function.

The second example (in dashed red) is the derivative of the SiLU activation function $\Phi'(x)=(1+e^{-x}+xe^{-x})/(1+e^{-x})^2$, with global $\arg\min$ and $\arg\max$ defined as $\underline{z}=-2.3994$ and $\overline{z}=2.3994$, respectively. \hfill $\triangle$
\end{example}

The only exceptions of activation functions whose derivative do not belong to this class that the author could find in the literature are: the binary step, which is discontinuous so its derivative is undefined in $0$; and the Gaussian activation function $\Phi(x)=e^{-x^2}$ which belongs to this class, but not its derivative (although $-\Phi'$ does belong to this class, so a very similar approach can still be applied).

\begin{figure}[tbp]
\centering
\includegraphics[width=\columnwidth]{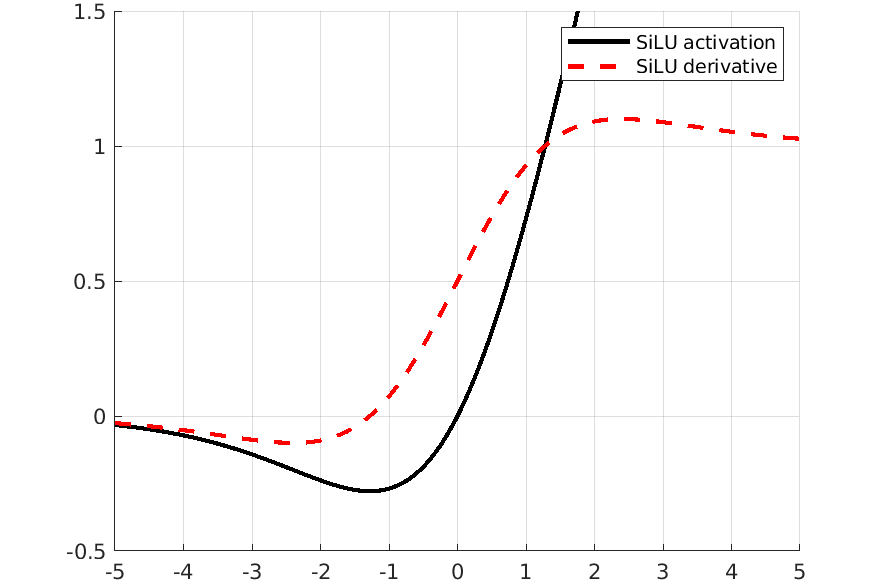}
\caption{SiLU activation function (black line) and its derivative (red dashed).}
\label{fig silu}
\end{figure}
\fi

\begin{proposition}
\label{prop local bounds}
Given a scalar function $\varphi$ as defined above and a bounded input domain $[\underline{x},\overline{x}]\in\R$, the local bounds of $\varphi$ on $[\underline{x},\overline{x}]$ are given by:
\begin{gather*}
\label{eq local bounds}
\min_{x\in[\underline{x},\overline{x}]}\varphi(x)=
\begin{cases}
\varphi(\underline{z}) &\text{if }\underline{z}\in[\underline{x},\overline{x}],\\
\min(\varphi(\underline{x}),\varphi(\overline{x}))&\text{otherwise},
\end{cases}
\\
\max_{x\in[\underline{x},\overline{x}]}\varphi(x)=
\begin{cases}
\varphi(\overline{z}) &\text{if }\overline{z}\in[\underline{x},\overline{x}],\\
\max(\varphi(\underline{x}),\varphi(\overline{x}))&\text{otherwise}.
\end{cases}
\end{gather*}
\end{proposition}

In this paper, Proposition~\ref{prop local bounds} is used to obtain local bounds of the activation function derivatives in order to compute bounds on the network's Jacobian matrix for Proposition~\ref{prop algebraic MM}.
However, Proposition~\ref{prop local bounds} can also be useful on activation functions themselves to compare (in Section~\ref{sec numerical}) our method with the naive interval propagation through the layers.

\section{Reachability algorithm}
\label{sec reachability}
Solving Problem~\ref{pb definition} by applying the mixed-monotonicity approach from Proposition~\ref{prop algebraic MM} on the neural network \eqref{eq ffnn} can be done in many different ways.
One possibility is to iteratively compute bounds on the Jacobian matrix of the network and then apply Proposition~\ref{prop algebraic MM} once for the whole network.
However, this may result in wide bounds for the Jacobian of the whole network, and thus in a fairly conservative over-approximation of the network output due to the term $\alpha^i$ in Proposition~\ref{prop algebraic MM}.
The dual approach is to iteratively apply Proposition~\ref{prop algebraic MM} to each layer of the network, which would result in tighter Jacobian bounds (since one layer's Jacobian requires less interval matrix products than the Jacobian of the whole network), and thus tighter over-approximation of the layer's output.
However, the loss of the dependency with respect to the network's input may induce another source of accumulated conservativeness if we have many layers.
Any intermediate approach can also be considered, where we split the network into several consecutive partial networks and apply Proposition~\ref{prop algebraic MM} iteratively to each of them.

Although all the above approaches result in over-approximations of the network output, we cannot determine in advance which method would yield the tightest bounds, since this is highly dependent on the considered network and input interval.
We thus devised an algorithm that encapsulates all possible choices by running Proposition~\ref{prop algebraic MM} on all partial networks within \eqref{eq ffnn}, before taking the intersection of the obtained bounds to tighten the bounds on each layer's output.
The main steps of this approach are presented in Algorithm~\ref{algo} and summarized below.

Given the $L$-layer network \eqref{eq ffnn} with activation function $\Phi$ and input interval $[\underline{x^0},\overline{x^0}]$ as in Problem~\ref{pb definition}, our goal is to apply Proposition~\ref{prop algebraic MM} to each partial network of \eqref{eq ffnn}, denoted as $\mathtt{NN}(k,l)$, containing only layers $k$ to $l$ (with $k\leq l$) and with input $x^{(k-1)}$ and output $x^{(l)}$.
We initialize the Jacobian bounds of each partial network to the identity matrix.
Then, we iteratively explore the network (going forward), where for each layer $l$ we first use interval arithmetics~\cite{jaulin2001applied} to compute the pre-activation bounds based on the knowledge of the output bounds of the previous layer, and then apply Proposition~\ref{prop local bounds} to obtain local bounds on the activation function derivative ($\varphi=\Phi'$).
Next, for each partial network $\mathtt{NN}(k,l)$ (with $k\leq l$) ending at the current layer $l$, we compute its Jacobian bounds based on the Jacobian of layer $l$ ($[\underline{\Phi'},\overline{\Phi'}]*W^{(l)}$) and the Jacobian of the partial network $\mathtt{NN}(k,l-1)$.
We then apply Proposition~\ref{prop algebraic MM} to the partial network $\mathtt{NN}(k,l)$ with the Jacobian bounds we just computed and input bounds $[\underline{x^{(k-1)}},\overline{x^{(k-1)}}]$.
Finally, once we computed the $l$ over-approximations of $x^{(l)}$ corresponding to each partial network ending at layer $l$, we take the intersection of all of them to obtain the final bounds for $x^{(l)}$.

\begin{algorithm2e}[tbh]
  \SetKwFunction{AlgebraicMM}{Prop2}
  \SetKwFunction{LocalBounds}{Prop3}
  \SetKwFunction{NN}{NN}
  \KwIn{$L$-layer network \eqref{eq ffnn}, input interval $[\underline{x^0},\overline{x^0}]$, activation function $\Phi$ (with $\Phi'$ defined as in Section~\ref{sub prelim af bounds})}
  $\forall k,l\in\{1,\dots,L\},~\underline{J(k,l)}\leftarrow I,\overline{J(k,l)}\leftarrow I$\\
  \For{$l\in\{1,\dots,L\}$}{
  $[\underline{\Phi'},\overline{\Phi'}]\leftarrow\LocalBounds(\Phi',W^{(l)}*[\underline{x^{(l-1)}},\overline{x^{(l-1)}}]+b^{(l)})$\\
    \For{$k\in\{1,\dots,l\}$}{
    $[\underline{J(k,l)},\overline{J(k,l)}]\leftarrow[\underline{\Phi'},\overline{\Phi'}]*W^{(l)}*[\underline{J(k,l-1)},\overline{J(k,l-1)}]$\\
    $[\underline{x(k,l)},\overline{x(k,l)}]\leftarrow\AlgebraicMM(\NN(k,l),[\underline{x^{(k-1)}},\overline{x^{(k-1)}}],[\underline{J(k,l)},\overline{J(k,l)}])$
    }
    $[\underline{x^{(l)}},\overline{x^{(l^)}}]\leftarrow[\underline{x(1,l)},\overline{x(1,l)}]\cap\dots\cap[\underline{x(l,l)},\overline{x(l,l)}]$
  }
  \KwOut{Over-approximation $[\underline{x^{(L)}},\overline{x^{(L)}}]$ of the network output}
\caption{Mixed-monotonicity reachability analysis of a feedforward neural network.\label{algo}}
\end{algorithm2e}

\begin{theorem}
\label{th reachability}
The interval $[\underline{x^{(L)}},\overline{x^{(L)}}]$ returned by Algorithm~\ref{algo} is a solution to Problem~\ref{eq ffnn}.
\end{theorem}

This is easily proved by the fact that we compute several interval over-approximations of the output of each layer using Proposition~\ref{prop algebraic MM}, which ensures that their intersection is still an over-approximation of the layer's output.
Then, using these over-approximations as the input bounds of the next layers guarantees the soundness of the approach.

Note also that in Algorithm~\ref{algo}, Proposition~\ref{prop algebraic MM} is applied to every partial network that exists within the main network \eqref{eq ffnn}.
Although this implies a computational complexity of $O(L^2)$, it guarantees that the resulting interval $[\underline{x^{(L)}},\overline{x^{(L)}}]$ from Algorithm~\ref{algo} is the least conservative solution to Problem~\ref{pb definition} that could be obtained from applying the mixed-monotonicity reachability analysis of Proposition~\ref{prop algebraic MM} to any decomposition of \eqref{eq ffnn} into consecutive partial networks.

\section{Numerical comparisons}
\label{sec numerical}
To evaluate the performances of our approach, we run Algorithm~\ref{algo} on both established benchmarks and randomly generated neural networks, and compare the results with state-of-the-art tools for reachability analysis and verification of neural networks.\,\footnote{All computations are done on a laptop with $1.80$GHz processor and $16$GB of RAM running Matlab 2021b.
The code used to generate all numerical comparisons described in this section is available at:\\ \url{https://gitlab.com/pj_meyer/MMRANN}}
Since our method is an optimization-free reachability analysis approach using interval bounds, we focus these numerical comparisons to the most relevant toolboxes of this category, and we leave to future work the additional comparisons with less related verification tools relying on optimization-based approaches~\cite{liu2019algorithms} or reachability analysis with more complex set representations (polyhedra, zonotopes) as in the ERAN~\cite{eran} and NNV libraries~\cite{tran2020nnv}.

We provide comparisons to the following $5$ tools and methods for reachability analysis of neural networks.
We first use as a baseline the naive interval bound propagation (IBP), as presented e.g.\ in~\cite{xiang2020reachable}, since this is the simplest approach using interval arithmetic at each layer, but it is also very conservative due to losing dependency to the network's inputs at each step.
Three of the other tools (\textsc{ReluVal}~\cite{shiqi2018reluval}, \textsc{Neurify}~\cite{shiqi2018neurify}, \textsc{VeriNet}~\cite{henriksen2020efficient}) rely on methods propagating symbolic intervals (i.e.\ bounding functions linear in $x^{(0)}$), sometimes combined with linear relaxations of the nonlinear activation functions.
The fifth tool (CROWN~\cite{zhang2018efficient}) instead relies on a backward propagation of these linear relaxations.

Since these tools are written in various programming languages (Matlab, C, Python), for the convenience of comparison we have re-implemented 
\ifarxiv
the bounding method of each tool in Matlab.
\fi
The implemented approaches are those described in each of the original papers cited above, and may thus slightly differ from the latest updates of the corresponding public toolboxes.
\ifarxiv
Note also that the comparisons in this section only focus on reachability analysis as formulated in Problem~\ref{pb definition} (over-approximating the network's reachable set for a given input uncertainty), and we thus disregard the iterative refinement algorithms that some of the above tools use to verify or falsify properties on the network's behavior.
\fi

\paragraph*{VNN benchmarks}
We first compare our mixed-monotonicity approach with the above $5$ reachability tools on the \emph{MNISTFC} benchmark as presented in the 2021 VNN competition~\cite{bak2021second}.
This benchmark is composed of $3$ feedfoward ReLU neural networks trained on the MNIST dataset for the recognition of handwritten digits.
All $3$ networks have $784$ inputs ($28\times 28$ pixels of a gray-scale picture), an output layer with $10$ nodes, and an additional $2$, $4$ or $6$ hidden layers of $256$ nodes each.
We run all $6$ reachability methods on these $3$ networks for $250$ different input pictures, each with an $\epsilon=0.03$ uncertainty around their nominal values.
In terms of tightness of the output bounds, our approach is $10$ to $1000$ tighter than IBP and has a comparable order of magnitude with the other four tools.
In the top half of Table~\ref{table vnn}, the first three columns summarize the percentage of the $250$ evaluated input pictures for which our mixed-monotonicity approach results in tighter (or at least as tight) output bounds than the other five methods.
We can see in particular that our method outperforms all others in at least $70\%$ cases on the benchmark with two hidden layers.
In terms of computation times (bottom of Table~\ref{table vnn}), our approach is significantly slower than IBP and \textsc{VeriNet} and a bit slower than CROWN, but it is also up to $3$ times faster than \textsc{ReluVal} and \textsc{Neurify}.

We ran the same comparison on the only non-ReLU feedforward benchmark of the 2021 VNN competition~\cite{bak2021second}: the \emph{ERAN-sigmoid} benchmark is also trained on the MNIST dataset and has the same structure as the above MNISTFC networks, but with $6$ hidden layers of $200$ nodes each, sigmoid activation functions and input uncertainty $\epsilon=0.012$.
On the same $250$ input pictures, we observe that: IBP is always more conservative than our approach (as expected); \textsc{ReluVal} and \textsc{Neurify} cannot be run on non-ReLU networks; \textsc{VeriNet} and CROWN both fail for every $250$ inputs because their results are so conservative on this benchmark that Matlab evaluates $e^{-x}=\infty$ when the pre-activation value $x$ goes too far in the negative, and the (theoretically horizontal) tangent of the sigmoid $e^{-x}/(1+e^{-x})^2$ cannot be computed.
\textsc{VeriNet} and CROWN's failure due to excessive conservativeness on this sigmoid benchmark is also observed when using an input uncertainty $\epsilon=1.2*10^{-5}$ which is $1000$ smaller than the one suggested in the VNN benchmark~\cite{bak2021second}.

\begin{table}[tbp]
\centering
\begin{tabular}{c | c c c c} 
{\bf Method} & {\bf mnistfc-$2$} & {\bf mnistfc-$4$} & {\bf mnistfc-$6$} & {\bf ERAN-sig}\\\hline
IBP~\cite{xiang2020reachable} & $100\%$ & $100\%$ & $100\%$ & $100\%$\\
\textsc{ReluVal}~\cite{shiqi2018reluval} & $80\%$ & $74.8\%$ & $65.6\%$ & - \\
\textsc{Neurify}~\cite{shiqi2018neurify} & $72.4\%$ & $72\%$ & $58.8\%$ & - \\
\textsc{VeriNet}~\cite{henriksen2020efficient} & $70.4\%$ & $54\%$ & $15.6\%$ & $100\%$ \\
CROWN~\cite{zhang2018efficient} & $70.4\%$ & $54\%$ & $15.6\%$ & $100\%$ \\
\hline
IBP~\cite{xiang2020reachable} & $0.012$ & $0.019$ & $0.025$ & $0.016$\\
\textsc{ReluVal}~\cite{shiqi2018reluval} & $42$ & $62$ & $85$ & - \\
\textsc{Neurify}~\cite{shiqi2018neurify} & $43$ & $62$ & $84$ & - \\
\textsc{VeriNet}~\cite{henriksen2020efficient} & $0.012$ & $0.019$ & $0.026$ & $0.017$ \\
CROWN~\cite{zhang2018efficient} & $0.78$ & $4.1$ & $10$ & $5.7$ \\
Mixed-Mono & $15$ & $36$ & $69$ &  $39$\\
\end{tabular}
\caption{Top: proportion of the $250$ MNIST input pictures for which our approach gives output bounds tighter than (or equal to) other methods. 
Bottom: average computation time in $s$.}
\label{table vnn}
\end{table}

\paragraph*{Random networks}
To get more representative comparisons on a wider variety of network dimensions (since the above benchmarks are only $4$ specific networks), and more importantly to highlight the main strength of our approach handling general activation functions (since we could not find well-established benchmarks using any activation other than ReLU and Sigmoid), we run these comparisons over two large sets of randomly generated neural networks.

We first consider a set of $10000$ small feedforward networks as defined in \eqref{eq ffnn} and whose parameters are randomly chosen as follows: depth $L\in\{1,\dots,5\}$; input and output dimensions $n_0,n_L\in\{1,\dots,10\}$; width of hidden layers (each layer may have a different width) $n_1,\dots,n_{L-1}\in\{1,\dots,30\}$.
The second set contains $1000$ deeper and wider networks with: depth $L\in\{5,\dots,10\}$; input dimension $n_0\in\{500,\dots,1000\}$; output dimension $n_L\in\{10,\dots,50\}$; width of hidden layers $n_1,\dots,n_{L-1}\in\{100,\dots,200\}$.
The input bounds as in Problem~\ref{pb definition} are defined as the hypercube $[\underline{x^{(0)}},\overline{x^{(0)}}]=\{x^*\}+[-0.1,0.1]^{n_0}$ around a randomly chosen center input $x^*\in[-1,1]^{n_0}$.
The comparison between our mixed-monotonicity method in Algorithm~\ref{algo} and the five approaches from the literature cited above (when applicable) is done for each of these $4$ activation functions:
\begin{itemize}
\item Rectified Linear Unit (ReLU) is the piecewise-affine function $\Phi(x)=\max(0,x)$, handled by all $6$ tools;
\item Hyperbolic tangent (TanH) is the S-shaped function $\Phi(x)=(e^x-e^{-x})/(e^x+e^{-x})$, handled by all tools apart from \textsc{ReluVal} and \textsc{Neurify};
\item Exponential Linear Unit (ELU) is the monotone function $\Phi(x)=e^x-1$ if $x\leq0$ and $\Phi(x)=x$ if $x\geq0$, natively handled only by IBP and our method.
ELU is compatible with the framework of \textsc{VeriNet} and CROWN (although not implemented) at the condition of providing a method to compute linear equations bounding $\Phi$ on any bounded input set.
We added this to our own Matlab implementation of these tools;
\item Sigmoid Linear Unit (SiLU) is the non-monotone function $\Phi(x)=x/(1+e^{-x})$, only natively handled by our method.
Although the implementation of IBP in~\cite{xiang2020reachable} is restricted to monotone activation functions, our own Matlab implementation of IBP is extended to handle non-monotone activation by using Proposition~\ref{prop local bounds}.
\end{itemize}
All these activation functions are Lipschitz continuous and their derivatives satisfy the desired shape described in Section~\ref{sub prelim af bounds}.
Algorithm~\ref{algo} can thus be applied to all of them.

\ifarxiv
For the set of $10000$ small networks, Table~\ref{table small time} gives the average computation times of each methods for each activation function.
For these time averaging to be meaningful despite the varying sizes of the networks, we actually divide the computation times by the total number of neurons in the considered network, and only then we take the average.
Table~\ref{table small tight equal} gives the proportion of the $10000$ small networks for which our mixed-monotonicity approach results in at least as tight output bounds as the other methods.
Table~\ref{table small tight strict} gives the same result but only when our approach is strictly tighter.
For the $1000$ large networks, the same results are displayed in Tables~\ref{table large time}-\ref{table large tight strict}.
Note that the time tables are in microseconds for small networks (Table~\ref{table small time}), and milliseconds for large ones (Table~\ref{table large time}).
In the next three paragraphs, we analyse these results and highlight the main take-away conclusions from these numerical comparisons in terms of generality of the methods, complexity, and tightness of the output bounds.
\fi

Among all $6$ compared methods, our mixed-monotonicity approach is the most general one and the only one that could natively handle all considered activation functions.
Indeed, \textsc{ReluVal} and \textsc{Neurify} are limited to piecewise-affine functions (such as ReLU) and \textsc{VeriNet} and CROWN cannot handle non-monotone functions (such as SiLU).
In addition, although we added an implementation for IBP to handle SiLU, and for \textsc{VeriNet} and CROWN to handle ELU, the original tools do not natively handle these activation functions.
\ifarxiv
(The corresponding results are given in parentheses in Tables~\ref{table small time}-\ref{table large tight strict}.)
\fi
This last comment brings up another important aspect of the generality of our mixed-monotonicity approach: the ease of implementation to add new activation functions.
Indeed, \textsc{Neurify}, \textsc{VeriNet} and CROWN rely on linear relaxations of the nonlinear activation functions, which may require long and complex implementations to be provided by the user for each new function (e.g.\ finding the optimal relaxations of sigmoid-shaped functions takes several hundreds of lines of code in the implementation of \textsc{VeriNet}).
In contrast, for a user to add a new activation type to be used within our mixed-monotonicity approach, all they need to provide is the definition of the activation function and its derivative, and the global $\arg\min$ and $\arg\max$ of the derivative as defined in Section~\ref{sub prelim af bounds}.
Everything else is automatically handled internally by Algorithm~\ref{algo}.

This generality and ease of use however comes at the cost of a higher computational complexity of $O(L^2)$ since Algorithm~\ref{algo} calls the mixed-monotonicity reachability method from Proportion~\ref{prop algebraic MM} on all $L(L+1)/2$ partial networks within the main $L$-layer network.
On the set of smaller networks, this implies that our approach is the slowest, with a similar order of magnitude as CROWN (which also has a complexity of $O(L^2)$), while the other four methods are 
\ifarxiv $10$ to $20$ times faster (Table~\ref{table small time}). \fi 
When the size of the network increases, IBP and \textsc{VeriNet} are always the fastest methods and keep a consistent average computation time (per number of neurons in the network), while all other methods see their computation time per neuron increase, the worst being \textsc{ReluVal} and \textsc{Neurify} which become slower than our mixed-monotonicity 
\ifarxiv approach (Table~\ref{table large time}). \fi 

Finally, in terms of tightness (measured as the $2$-norm width of the output bounds), we first observe that, as expected, our approach is always at least as tight as the naive IBP 
\ifarxiv method (Tables~\ref{table small tight equal} and~\ref{table large tight equal}), \fi 
and even strictly tighter in $70$-$80\%$ cases in the set of small 
\ifarxiv networks (Table~\ref{table small tight strict}). \fi 
Apart from the very conservative IBP, it should be noted that none of the other $5$ methods always outperforms the others.
For example on the set of $10000$ small ReLU networks, our approach is strictly tighter than the other four in $12$-$36\%$ cases, equal in $31\%$ cases, and strictly looser in the remaining $32$-$57\%$ 
\ifarxiv cases (Tables~\ref{table small tight equal} and~\ref{table small tight strict}). \fi 
The respective strengths of each method thus make them complementary.
On average, our approach returns at least as tight output bounds as the state-of-the-art tools in less than half cases on smaller networks, but this percentage significantly increases on larger networks 
\ifarxiv except for ReLU (Tables~\ref{table small tight equal} and~\ref{table large tight equal}). \fi 
When the size of the network increases, we also observe a larger proportion of cases where the compared methods return approximately equal output bounds, particularly on activation functions with approximate saturations, such as 
\ifarxiv TanH (Tables~\ref{table large tight equal} and~\ref{table large tight strict}). \fi 
However, one interesting case if with the ELU activation (for which the state-of-the-art tools where not designed and optimized), where our approach not only outperforms both \textsc{VeriNet} and CROWN on almost all $1000$ tested large networks, but it also obtains strictly tighter output bounds than \textsc{VeriNet} in $79\%$ cases, and CROWN is seen to fail to return a numerical result in all cases due to its excessive conservativeness and its formulation in~\cite{zhang2018efficient} being incompatible with horizontal linear relaxations (similarly to the ERAN-sigmoid benchmark).

\ifarxiv

\begin{table}[tbp]
\centering
\begin{tabular}{c | c c c c} 
{\bf Method} & {\bf ReLU} & {\bf TanH} & {\bf ELU} & {\bf SiLU}\\\hline
IBP~\cite{xiang2020reachable} & $12$ & $18$ & $11$ & ($13$)\\
\textsc{ReluVal}~\cite{shiqi2018reluval} & $29$ & - & - & - \\
\textsc{Neurify}~\cite{shiqi2018neurify} & $27$ & - & - & - \\
\textsc{VeriNet}~\cite{henriksen2020efficient} & $14$ & $33$ & ($25$) & - \\
CROWN~\cite{zhang2018efficient} & $199$ & $213$ & ($177$) & - \\
Mixed-Monotonicity & $591$ & $462$ & $550$ &  $543$\\
\end{tabular}
\caption{Average computation time (per neuron in the network) in $\mu s$ for the set of $10000$ small networks.}
\label{table small time}
\end{table}

\begin{table}[tbp]
\centering
\begin{tabular}{c | c c c c} 
{\bf Method} & {\bf ReLU} & {\bf TanH} & {\bf ELU} & {\bf SiLU}\\\hline
IBP~\cite{xiang2020reachable} & $100\%$ & $100\%$ & $100\%$ & ($100\%$)\\
\textsc{ReluVal}~\cite{shiqi2018reluval} & $68\%$ & - & - & - \\
\textsc{Neurify}~\cite{shiqi2018neurify} & $46\%$ & - & - & - \\
\textsc{VeriNet}~\cite{henriksen2020efficient} & $43\%$ & $32\%$ & ($40\%$) & - \\
CROWN~\cite{zhang2018efficient} & $43\%$ & $31\%$ & ($38\%$) & - \\
\end{tabular}
\caption{Proportion of the $10000$ small random networks for which our Mixed-Monotonicity approach results in tighter (or equal) output bounds than other methods.}
\label{table small tight equal}
\end{table}

\begin{table}[tbp]
\centering
\begin{tabular}{c | c c c c} 
{\bf Method} & {\bf ReLU} & {\bf TanH} & {\bf ELU} & {\bf SiLU}\\\hline
IBP~\cite{xiang2020reachable} & $73\%$ & $71\%$ & $79\%$ & ($79\%$)\\
\textsc{ReluVal}~\cite{shiqi2018reluval} & $36\%$ & - & - & - \\
\textsc{Neurify}~\cite{shiqi2018neurify} & $16\%$ & - & - & - \\
\textsc{VeriNet}~\cite{henriksen2020efficient} & $12\%$ & $11\%$ & ($19\%$) & - \\
CROWN~\cite{zhang2018efficient} & $12\%$ & $10\%$ & ($17\%$) & - \\
\end{tabular}
\caption{Proportion of the $10000$ small random networks for which our Mixed-Monotonicity approach results in strictly tighter output bounds than other methods.}
\label{table small tight strict}
\end{table}


\begin{table}[tbp]
\centering
\begin{tabular}{c | c c c c} 
{\bf Method} & {\bf ReLU} & {\bf TanH} & {\bf ELU} & {\bf SiLU}\\\hline
IBP~\cite{xiang2020reachable} & $0.016$ & $0.018$ & $0.016$ & ($0.018$)\\
\textsc{ReluVal}~\cite{shiqi2018reluval} & $44$ & - & - & - \\
\textsc{Neurify}~\cite{shiqi2018neurify} & $44$ & - & - & - \\
\textsc{VeriNet}~\cite{henriksen2020efficient} & $0.034$ & $0.05$ & ($0.037$) & - \\
CROWN~\cite{zhang2018efficient} & $4.8$ & $4.8$ & ($4.8$) & - \\
Mixed-Monotonicity & $33$ & $29$ & $34$ &  $33$\\
\end{tabular}
\caption{Average computation time (per neuron in the network) in $ms$ for the set of $1000$ large networks.}
\label{table large time}
\end{table}

\begin{table}[tbp]
\centering
\begin{tabular}{c | c c c c} 
{\bf Method} & {\bf ReLU} & {\bf TanH} & {\bf ELU} & {\bf SiLU}\\\hline
IBP~\cite{xiang2020reachable} & $100\%$ & $100\%$ & $100\%$ & ($100\%$)\\
\textsc{ReluVal}~\cite{shiqi2018reluval} & $100\%$ & - & - & - \\
\textsc{Neurify}~\cite{shiqi2018neurify} & $2\%$ & - & - & - \\
\textsc{VeriNet}~\cite{henriksen2020efficient} & $0.1\%$ & $100\%$ & ($98\%$) & - \\
CROWN~\cite{zhang2018efficient} & $0.1\%$ & $100\%$ & ($100\%$) & - \\
\end{tabular}
\caption{Proportion of the $1000$ large random networks for which our Mixed-Monotonicity approach results in tighter (or equal) output bounds than other methods.}
\label{table large tight equal}
\end{table}

\begin{table}[tbp]
\centering
\begin{tabular}{c | c c c c} 
{\bf Method} & {\bf ReLU} & {\bf TanH} & {\bf ELU} & {\bf SiLU}\\\hline
IBP~\cite{xiang2020reachable} & $0\%$ & $0\%$ & $0\%$ & ($0\%$)\\
\textsc{ReluVal}~\cite{shiqi2018reluval} & $0\%$ & - & - & - \\
\textsc{Neurify}~\cite{shiqi2018neurify} & $0\%$ & - & - & - \\
\textsc{VeriNet}~\cite{henriksen2020efficient} & $0\%$ & $0\%$ & ($79\%$) & - \\
CROWN~\cite{zhang2018efficient} & $0\%$ & $2\%$ & ($100\%$) & - \\
\end{tabular}
\caption{Proportion of the $1000$ large random networks for which our Mixed-Monotonicity approach results in strictly tighter output bounds than other methods.}
\label{table large tight strict}
\end{table}

\fi

\section{Conclusions}
\label{sec conclu}
This paper presents a new method for the sound reachability analysis of feedforward neural networks using mixed-monotonicity.
The main strength of our approach is its very broad applicability to any network with Lipschitz-continuous activation functions, which is satisfied by the vast majority of activation functions currently in use in the literature.
Another significant strength of our framework is the greater simplicity to implement new activation functions compared to existing tools relying on linear relaxations.
We compared our approach with the naive Interval Bound Propagation and $4$ state-of-the-art tools using interval reachability analysis on neural network (\textsc{ReluVal}, \textsc{Neurify}, \textsc{VeriNet}, CROWN) on both benchmarks from the VNN competition~\cite{bak2021second} and sets of randomly generated neural networks of a wide variety of depth and width.
From these comparisons, we observed a good complementarity between our approach and the state-of-the-art tools (when applicable on the considered activation functions), where each outperforms the others on a significant percentage of the tested inputs and networks.

Although this tool can already be used on its own for the reachability analysis of neural networks, our next objective is to address its main limitation of a $O(L^2)$ complexity to make it more compatible with iterative refinement approaches similar to existing neural network verification algorithms.


\bibliographystyle{IEEEtran}
\bibliography{RA_of_NN_using_MM}

\end{document}